\documentclass[a4paper,fleqn]{cas-sc}

\usepackage[numbers]{natbib}
\usepackage{subfig}
\usepackage{url}            % simple URL typesetting
\usepackage{booktabs}       % professional-quality tables
\usepackage{amsfonts}       % blackboard math symbols
\usepackage{nicefrac}       % compact symbols for 1/2, etc.
\usepackage{microtype}      % microtypography
\usepackage{amsmath}
\usepackage{graphicx}
\usepackage{multirow}
\usepackage{adjustbox}
\usepackage{pifont}% http://ctan.org/pkg/pifont
\usepackage{cleveref}
\usepackage{todonotes}
\usepackage{commath}
\usepackage{mathtools}
\usepackage{amsthm}
\usepackage{amssymb}
\usepackage{tikz}
\usepackage{xcolor}
\usepackage{hyperref}
\usepackage{balance}  % to better equalize the last page
\usepackage{times}    % comment if you want LaTeX's default font
\usepackage{soul}
\usepackage{color}
\newcommand{\eg}{\textit{e.g\@.}}
\newcommand{\ie}{\textit{i.e\@.}}

\definecolor{LightCyan}{rgb}{0.83,0.89,0.75}
\usepackage{adjustbox}

\usepackage{doi}
\usepackage[ruled,linesnumbered]{algorithm2e}

\SetKwInput{KwInput}{Input}                % Set the Input
\SetKwInput{KwOutput}{Output}              % set the Output
\SetKwRepeat{Do}{do}{while} % for dowhile
\usepackage{mathtools, nccmath}
\usepackage{wrapfig}
\usepackage{comment}
\usepackage{graphicx}
\usepackage{xspace}
\usepackage{adjustbox}
\usepackage{lipsum}
\usepackage{tcolorbox}
\newtcolorbox{mybox}{colback=black!5!white,colframe=black,bottomrule=.25mm,toprule=.25mm,leftrule=.25mm,rightrule=.25mm,left=.25mm,right=.25mm,top=-.25mm,bottom=-.25mm}
\def\tsc#1{\csdef{#1}{\textsc{\lowercase{#1}}\xspace}}
\tsc{WGM}
\tsc{QE}
\tsc{EP}
\tsc{PMS}
\tsc{BEC}
\tsc{DE}
\newcommand{\paragraphb}[1]{\noindent{\bf #1} }

\begin{document}
\let\WriteBookmarks\relax
\def\floatpagepagefraction{1}
\def\textpagefraction{.001}

%\shorttitle{Leveraging social media news}

%\shortauthors{}

\title [mode = title]{Correction: Harnessing Federated Generative Learning for Green and Sustainable Internet of Things}                      

\affiliation[1]{organization={School of Computer Science, University of Electronic Science and Technology of China, Zhongshan Institute, Zhongshan 528402},
country={China}
    % citysep={}, % Uncomment if no comma needed between city and postcode
    }
\author[1]{Yuanhang Qi}[style=chinese]
\ead{qiyuanhang@zsc.edu.cn}
\credit{Conceptualization of this study, Methodology, Software}

\author[2]{M. Shamim Hossain}\cormark[1]

\ead{mshossain@ksu.edu.sa}
%\ead[URL]{www.sayahna.org}

\credit{Data curation, Writing - Original draft preparation}

\affiliation[2]{organization={Research Pervasive and Mobile Computing, and Department of Software Engineering, College of Computer and Information Sciences, King Saud University, Riyadh 13272},
   country={Saudi Arabia}}
  
\cortext[cor1]{Corresponding author}

%\nonumnote{This note has no numbers. In this work we demonstrate $a_b$
%  the formation Y\_1 of a new type of polariton on the interface
%  between a cuprous oxide slab and a polystyrene micro-sphere placed
%  on the slab.
%  }

\begin{abstract}
The rapid proliferation of devices in the Internet of Things (IoT) has ushered in a transformative era of data-driven connectivity across various domains. However, this exponential growth has raised pressing concerns about environmental sustainability and data privacy. In response to these challenges, this paper introduces One-shot Federated Learning (OSFL), an innovative paradigm that harmonizes sustainability and machine learning within IoT ecosystems. OSFL revolutionizes the traditional Federated Learning (FL) workflow by condensing multiple iterative communication rounds into a single operation, thus significantly reducing energy consumption, communication overhead, and latency. This breakthrough is coupled with the strategic integration of generative learning techniques, ensuring robust data privacy while promoting efficient knowledge sharing among IoT devices. By curtailing resource utilization, OSFL aligns seamlessly with the vision of green and sustainable IoT, effectively extending device lifespans and mitigating their environmental footprint. Our research underscores the transformative potential of OSFL, poised to reshape the landscape of IoT applications across domains such as energy-efficient smart cities and groundbreaking healthcare solutions. This contribution marks a pivotal step towards a more responsible, sustainable, and technologically advanced future.
\end{abstract}

% \includegraphics{figs/grabs.pdf}

%\item We customize a new one-shot federated generative learning communication network architecture for green and sustainable IoT.
%\item We propose two new local prompt generation strategies to greatly reduce communication overhead and latency.
%\item We develop a one-shot FL method with generative learning as the core, which is organically combined with generation to improve the overall performance of FL. 

% Keywords

\maketitle

\section{Introduction}
\textcolor{black}{The advent of the Internet of Things (IoT)~\cite{ref-1,ref-2,ref-3} has ushered in a new era of connectivity and data-driven decision-making across various domains~\cite{liu2020deep,liu2020privacy,liu2022right,du2023beyond}.} IoT devices have permeated our daily lives, enabling a wide range of applications, from smart homes and cities to industrial automation and healthcare. However, the proliferation of these devices has raised concerns about their environmental impact, especially with regard to energy consumption and data privacy. In this context, Federated Learning (FL)~\cite{mcmahan2017communication} has emerged as a promising paradigm to address these challenges by enabling collaborative machine learning on decentralized IoT devices, while also respecting user privacy~\cite{ref-4,ref-5,ref-6}.

Federated learning is a decentralized machine learning approach that enables multiple devices to collaboratively train a global model while keeping their data localized and private~\cite{li2020federated}. The workflow begins with the initialization of a global model, followed by device participation, where decentralized clients perform local training on their data, compute model updates, and securely aggregate them on a central server~\cite{bonawitz2019towards,guo2023promptfl}. This iterative process refines the global model over multiple rounds, monitored by evaluation metrics until the desired performance is achieved. \textcolor{black}{FL's privacy-preserving and distributed nature makes it suitable for applications like green and sustainable IoT, healthcare, and federated edge computing, addressing data privacy concerns while enabling machine learning in diverse and resource-constrained environments~\cite{zhou2021two,alagheband2022advanced,hajian2022secure,mohammed2023homomorphic,peng2023fedgm}.}

Traditional FL approaches often require numerous communication rounds between the central server and participating devices, which can result in significant energy consumption, latency, and communication overhead~\cite{kang2020scalable,kang2022communication}. {Inspired by previous work~~\cite{zhang2022dense,salehkaleybar2021one,heinbaugh2022data,zhang2023federated}, this paper introduces a novel approach known as ``One-shot Federated Learning(OSFL),'' aimed at minimizing resource utilization while maximizing the sustainability of IoT networks.} Leveraging the power of generative learning~\cite{zhang2023federated}, this approach allows IoT devices to learn and share knowledge efficiently in a single communication round, reducing the need for constant connectivity and data transmission. This approach not only promotes green and sustainable IoT but also enhances the scalability and practicality of FL in resource-constrained environments.

The need for sustainable IoT solutions has become increasingly pressing as the world grapples with the challenges of climate change and resource depletion~\cite{khaleel2023fault}. IoT devices are projected to proliferate exponentially in the coming years, making it imperative to design and adopt technologies that can mitigate their environmental impact~\cite{sabovic2023towards}. By incorporating generative learning techniques into the FL framework, this paper presents a promising avenue for reducing energy consumption and prolonging the lifespan of IoT devices, thereby contributing to the broader goal of sustainable technology development~\cite{triastcyn2020federated}.

Privacy concerns also loom large in the IoT landscape, as the data generated by these devices often contains sensitive information~\cite{alagheband2022advanced}. Traditional FL methods inherently preserve user privacy by keeping data localized, but they still involve multiple interactions that raise potential security risks~\cite{wei2020federated,li2021privacy}. OSFL enhances privacy further by minimizing data exposure and limiting communication, making it an attractive choice for applications where privacy is paramount. This paper explores how this approach can strike a delicate balance between data utility, model performance, and user privacy in green and sustainable IoT environments. To this end, we summarize the challenges we face as follows:

\begin{itemize}
    \item \textbf{(C1.) Difficult to Converge:} \textcolor{black}{ Since OSFL only communicates once, this brings severe challenges to global model convergence. Unlike traditional FL, OSFL cannot iteratively update the model and perform multiple rounds of communication~\cite{zhang2023federated}.}
    \item  \textbf{(C2.) New Privacy Concerns:}  \textcolor{black}{Although OSFL only conducts one-shot communication to mitigate privacy risks to a certain extent, adversaries still have the opportunity to threaten OSFL's privacy. The main reason is that the knowledge uploaded by OSFL is not protected by privacy protection measures.}
\end{itemize}

\textcolor{black}{To tackle the aforementioned challenges, this paper utilizes the concept of OSFL as a novel and environmentally friendly approach to machine learning within the realm of the IoT. {Specifically, we present an innovative framework termed Federated Generative Learning (FGL) which was proposed by Zhang et al.~\cite{zhang2023federated}.} FGL harnesses the robust capabilities of generative models, including Stable Diffusion and GPT-4V, to generate high-fidelity surrogate training data directly on the server. This process is guided by prompts provided by individual clients. In a concrete sense, our method streamlines the workflow by having each client upload only their respective prompts, linked to their local training data. Once all prompts are gathered from the clients, the server orchestrates prompt aggregation and subsequently synthesizes a premium-quality surrogate training dataset. This dataset can then be employed to train a global model.} By leveraging generative learning techniques and minimizing communication rounds, this approach aims to address the challenges of energy efficiency, scalability, and privacy in IoT networks. \textcolor{black}{The subsequent sections of this paper will delve into the technical details, experiments, and practical implementations of OSFL, illustrating its potential to usher in a greener and more sustainable era for IoT applications.} This paper's primary contributions are as follows:

\begin{itemize}
    \item We customize a new one-shot federated generative learning system for green and sustainable IoT, which can effectively alleviate data privacy and communication overhead issues.

    \item  We propose two new local prompt generation strategies, which aim to extract fine-grained feature information of local data to enhance the model's generalization ability.
    
    \item 
We develop a one-shot FL method with generative learning as the core, which is organically combined with generation to improve the overall performance of FL. In particular, the one-shot FL method based on generative learning significantly improves the problem of expensive communication in IoT.
    
    \item
  We conduct comprehensive validation and evaluation on three benchmark datasets, and the experimental results demonstrate the effectiveness of the proposed method.

\end{itemize}

The paper's structure is organized as follows: In Section \ref{sec-2}, we provide a comprehensive overview of the background and concepts related to FL and the IoT. This section highlights the existing challenges and open problems within this context. Section \ref{sec-3} delves into the foundational aspects of federated learning and text-to-image models, offering a clear understanding of these crucial components. Furthermore, we explore the practical application scenarios of text-to-image models and delve into the core challenges currently faced in this field. In Section \ref{sec-4}, we introduce our innovative One-shot FGL system, detailing its architecture and mechanisms. We then proceed to Section \ref{sec-5}, where we present a series of case studies conducted on three real-world datasets. In these case studies, we thoroughly analyze the experimental results, providing valuable insights and observations. Lastly, in Section \ref{sec-6}, we draw our conclusions, summarizing the key findings and contributions of our work in the context of OSFL for green and sustainable IoT. We summarize the notation descriptions used in this paper in Table \ref{tab-1}.

\begin{table}[!t]
    \centering
    \caption{{Glossary of symbols.}}
    \label{tab-1}
    \resizebox{0.5\linewidth}{!}{
    \begin{tabular}{|c|c|}
        \hline
        $B$ & The mini-batch size\\
        \hline
        $\mathcal{B}$ & The training batch set\\
        \hline
        $\mathcal{D}^{t}_{k}$ & The local training data at training round $t$ \\
        \hline
        $n_k$ & The number of training samples at $k$-th client\\
        \hline
        $\mathcal{D}_s$ & The task-agnostic surrogate dataset \\
        \hline
        $K$ & The number of clients\\
        \hline
        $\mathcal{K}$ & The client set\\
        \hline
        $p$ & The prompt generated by edge devices\\
        \hline
        $\mathcal{S}$ &The Synthetic training dataset\\
        \hline
        $\omega^{t}$ & The global model at training round $t$\\
        \hline
        $\ell _k$ & The local training loss\\
                \hline
        $\ell _{gen}$ & The generator loss\\
                \hline
        $\ell _{disc}$ & The discriminator loss\\
        \hline
        $G_t(\cdot.\cdot)$ & The denoising network\\
        \hline
    \end{tabular}
    }
\end{table}

\section{Related Work}\label{sec-2}
In the pursuit of advancing green and sustainable IoT through the lens of FL, this section provides an overview of existing research in four key subsections:

\subsection{Federated Learning in IoT}
\textcolor{black}{The integration of FL in the IoT has garnered substantial attention, primarily for its ability to harmonize the divergent requirements of data-driven intelligence and sustainability in resource-constrained environments. Prior research in this domain has delved into diverse FL strategies, encompassing federated optimization techniques, communication-efficient algorithms, and privacy-preserving methods.} Notable contributions include the works of McMahan et al. \cite{mcmahan2017communication} on federated averaging, which laid the foundation for distributed model training, and Bonawitz et al. \cite{bonawitz2016practical} on secure aggregation, addressing privacy concerns in FL. Moreover, recent research by Wei et al. \cite{wei2020federated} introduced the concept of "Federated Learning with Differential Privacy," further enhancing privacy guarantees in IoT applications. We build upon these foundations and extend them by introducing One-shot Federated Learning, which minimizes communication rounds and maximizes sustainability through generative learning.

Efforts to minimize communication between IoT devices and the central server have been pivotal in the context of FL for green IoT~\cite{sattler2019robust}. Various federated optimization~\cite{li2020federated,pathak2020fedsplit,wang2020tackling} techniques have been proposed to reduce the amount of information exchanged during each communication round. Notable works include the FedPAQ algorithm introduced by \cite{reisizadeh2020fedpaq}, which enables IoT devices to send only model updates to the central server rather than transmitting raw data. This approach significantly reduces communication bandwidth, making it an essential step towards greener and more sustainable IoT practices.

Quantization and compression techniques have emerged as effective means to further enhance communication efficiency in FL for IoT~\cite{chen2021communication,mcmahan2017communication}. Researchers have explored methods to quantize model updates into compact representations, thereby reducing the data transfer requirements between clients and the central server. Chen et al. \cite{chen2021communication} proposed a quantization approach, Q-FFL, which compresses model updates to a fraction of their original size, alleviating the communication burden on IoT devices. Such techniques align with the green IoT agenda by conserving energy and reducing the carbon footprint associated with data transmission.

\subsection{Sustainability in IoT}
Promoting environmental sustainability in IoT has become a critical research area, given the ever-growing number of IoT devices and their potential ecological impact~\cite{bibri2018iot,ali2021towards}. Previous studies have examined energy-efficient communication protocols, low-power hardware design, and green data center technologies. Additionally, researchers have proposed solutions to optimize IoT device lifespans and reduce their carbon footprint. The work of Xue et al. \cite{xue2023hybrid} on energy-efficient IoT communication protocols and Zhang et al. \cite{zhang2022artificial} on sustainable IoT design are notable examples. 

Additionally, advancements in green data center technologies and eco-friendly IoT infrastructure have contributed to reducing the overall carbon footprint of IoT deployments. These research endeavors underscore the importance of mitigating environmental impact, a key objective that aligns closely with the vision of our proposed One-shot Federated Learning approach, which seeks to enhance sustainability in IoT by optimizing machine learning while minimizing energy consumption and communication overhead. Our paper contributes to this body of knowledge by presenting One-shot Federated Learning as a novel approach that minimizes energy consumption during model updates and enhances the sustainability of IoT networks.

\subsection{One-shot Learning in FL}
The hallmark of OSFL~\cite{zhang2022dense,salehkaleybar2021one,chiang2023optimal,zhang2023federated} is its ability to streamline the FL process into a single communication round, minimizing the energy expended on data transmission and device connectivity. Traditional FL workflows require multiple rounds of communication to exchange model updates and converge towards a global model~\cite{mcmahan2017communication}. In contrast, OSFL condenses this multi-round process into a one-shot operation, reducing the overall energy footprint of IoT devices. This reduction in energy consumption is of paramount importance in green and sustainable IoT, where power-efficient operations are essential to mitigate environmental impact and prolong the lifespan of battery-operated devices.

The introduction of OSFL underscores the importance of striking a balance between sustainability and model performance in IoT applications. While reducing energy consumption and communication overhead are critical sustainability objectives, it is equally crucial to ensure that model quality and learning performance are not compromised. OSFL addresses this challenge by optimizing generative learning processes to create high-quality synthetic updates that faithfully represent the knowledge of participating devices~\cite{jhunjhunwala2023towards}. Through this balance, OSFL presents a compelling solution for green and sustainable IoT, enabling energy-efficient machine learning while maintaining or even improving the quality of learned models. Recent research has further pushed the boundaries of FL efficiency by exploring one-shot learning techniques. Li et al. \cite{li2020practical} demonstrated the feasibility of one-shot learning, enabling models to generalize from limited data instances. While these advancements primarily targeted traditional FL, they underscored the potential for a more streamlined and efficient FL workflow. Our work bridges the gap between communication-efficient FL and one-shot learning, introducing OSFL as a novel approach that synergizes these concepts to maximize sustainability in IoT while minimizing the environmental footprint of machine learning operations.

\subsection{Generative Learning in FL}
Generative learning has emerged as a powerful technique in machine learning, enabling models to generate data samples that capture underlying data distributions~\cite{creswell2018generative}. {In the context of FL, generative learning has been applied to enhance data privacy, as seen in the work of Nasr et al. \cite{nasr2019comprehensive} and Zhang et al.~\cite{zhang2023federated} on generative models, \eg, generated adversarial networks (GANs) and Stable Diffusion, for privacy-preserving data sharing.} Furthermore, recent research by Zhang et al. \cite{zhang2023federated} demonstrated the feasibility of generative models for one-shot learning tasks. 

Generative Learning has also played a role in data augmentation for FL. Techniques such as federated data generation, as proposed by Xin et al. \cite{xin2022federated}, leverage generative models to create synthetic data samples that help balance class distribution and improve model robustness in FL scenarios with imbalanced data. Furthermore, the application of generative models in one-shot learning tasks has demonstrated significant promise. Zhang et al. \cite{zhang2023towards} explored the use of generative models for efficient adaptation to new tasks, enabling FL models to learn from a single example. This concept has implications for resource-efficient machine learning and can be integrated into the FL framework to optimize knowledge transfer with minimal communication.

In the context of our paper, we leverage the capabilities of generative learning within the FL workflow to introduce OSFL, a novel approach that combines generative learning and FL to maximize sustainability in IoT while minimizing communication overhead and energy consumption. This integration represents a pivotal advancement in the field, addressing the pressing need for eco-friendly and efficient machine learning operations in IoT environments. Building upon these foundations, our paper leverages generative learning to optimize the FL workflow, reducing communication overhead and fostering sustainable IoT by achieving learning objectives with minimal data transfer and device energy consumption.

\section{Preliminaries}\label{sec-3}
\subsection{Federated Learning}
FL empowers decentralized participants to engage in collaborative machine learning model training without the need to share their private data with one another, as highlighted in \cite{mcmahan2017communication}. In a conventional supervised FL framework, the structure typically comprises a server denoted as $S$ and a set of $K$ clients. Each client maintains a comprehensive labeled dataset represented as $\mathcal{D}$, while the server itself does not possess any data. This architecture ensures data privacy and security during the collaborative training process.
% While ensuring that the raw data is not shared between clients, the goal of FL is to aggregate the local models trained by all clients to obtain a global model with almost equivalent performance to centralized learning.
Specifically, a widely used FL training strategy (i.e. FedAvg \cite{mcmahan2017communication}) is described as follows.

At the beginning of FL, the server initializes an ML model parameter $\omega_{0}$ and broadcasts it to a subset of clients (i.e., $\mathcal{K}_0$, and $\mathcal{K}_0 \subseteq \mathcal{K}$) for local training. Afterward, for each communication round $t$, the clients in the selected subset $\mathcal{K}_t$ train the received global model $\omega_{t}$ on the local dataset. For instance, the minimization objective function for the $k$-th client training a local model $\omega_{t}^{k}$ (i.e., $\omega_{t}^{k} \gets \omega_{t}$) on dataset $\mathcal{D}_k=\{(x_1, y_1), \cdots ,(x_{n_k}, y_{n_k})\}$ can be formulated as follows.
\begin{equation}
{\ell _k} = \frac{1}{{{n_k}}}\sum\limits_{i = 1}^{{n_k}} {{L_i}({y_i},p(f({x_i};{\omega ^k}))}.
\end{equation}
In this context, we define the variables as follows: $n_{k}$ represents the number of training samples held by client $k$, $\ell_{i}$ represents the loss function applied to the $i$-th training sample, $y_{i}$ denotes the ground-truth corresponding to $x_{i}$, and $p(y|x_{i};\omega_{t}^{k})$ signifies the probability vector predicted by model $\omega_{t}^{k}$ for the $i$-th training sample. Concurrently, model $\omega_{t}^{k}$ undergoes continuous updates until the completion of the local training process by client $k$.
\begin{equation}
	\omega_{t+1}^{k} = \omega_{t}^{k} - \eta\bigtriangledown \mathcal{L}_{k}(\mathcal{D}_k; \omega_{t}^{k}).
\end{equation}
In this context, $\eta$ represents the learning rate, governing the magnitude of each model update step. Subsequently, within the subset $\mathcal{K}_t$, each client uploads their locally trained model to the server for aggregation, with the aggregation formula outlined as follows.
\begin{equation}\label{eq-3}
    \omega_{t+1} = \sum\nolimits_{k \in {\mathcal{K}_t}} {p_k} \omega_{t+1}^{k},
\end{equation}
where $p_k=\frac{n_k}{\sum\nolimits_{k\in {\mathcal{K}_t}}n_k}$ represents the contribution rate of the $k$-th client to the current global model. Again, the server selects a subset of clients for the next communication round to send down the global model $\omega_{t+1}$ until the termination condition is reached.

\subsection{Text-to-Image Generative Models}
\textcolor{black}{Text-to-Image Generative Models (T2IGMs) \cite{ding2021cogview,liu2020deep,bird2023typology} signify a groundbreaking advancement at the convergence of natural language processing and computer vision. These models are specifically engineered to produce visually coherent and contextually relevant images based on textual descriptions, effectively closing the traditional gap between linguistic and visual comprehension. In recent years, T2IGMs have witnessed notable advancements, emerging as a formidable tool with applications ranging from content creation to augmenting the capabilities of IoT systems \cite{deckers2023infinite,ferdowsi2019generative}.} This section provides a comprehensive background and definition of T2IGMs, tracing their evolution and elucidating their profound impact across diverse domains.

T2IGMs typically consist of two primary components: an encoder-decoder architecture and a generative model~\cite{ding2021cogview}. The encoder processes the textual input, converting it into a latent representation that captures the semantics of the text. The decoder, on the other hand, takes this latent representation and generates a corresponding image. Often, the generative model incorporates GAN-based techniques to ensure the generated images are both visually convincing and semantically faithful to the input text~\cite{goodfellow2020generative,goodfellow2014generative}. The synergy between these components enables T2IGMs to understand and translate textual descriptions into high-quality images. The training of a T2IGM involves a GAN-like setup with a discriminator and generator~\cite{liu2019ppgan}. The discriminator tries to distinguish between real images and images generated from text, while the generator aims to generate images that fool the discriminator. The loss function typically involves both a generator loss ${\ell _{gen}}$ and a discriminator loss ${\ell _{disc}}$:

\begin{equation}
\begin{aligned}
& \mathcal{L}_{\text {gen }}=-\log (D(\text { Generated Image })) \\
& \mathcal{L}_{\text {disc }}=-\log (D(\text { Real Image }))-\log (1-D(\text { Generated Image })),
\end{aligned}
\end{equation} 
where $D$ represents the discriminator network. The overall training objective is to find the parameters of the generator network that minimize the generator loss while simultaneously minimizing the discriminator loss. This is often represented as a min-max optimization problem:
\begin{equation}
\min _{f_{\text {generator }}} \max _{f_{\text {discriminator }}}\left[\mathcal{L}_{\text {gen }}-\lambda \mathcal{L}_{\text {disc }}\right],
\end{equation}
where $\lambda$ is a hyperparameter that controls the trade-off between the generator and discriminator losses.

The versatility of T2IGMs has led to their adoption across a wide spectrum of domains. In the realm of art and design, T2IGMs serve as creative assistants, turning textual prompts into stunning visual artworks. In e-commerce, they facilitate the automatic generation of product images from textual descriptions, streamlining the process of catalog creation~\cite{fan2022automatic,yang2016automatic}. For accessibility, T2IGMs have been utilized to provide visually impaired individuals with rich, textual descriptions converted into tactile and comprehensible images. Additionally, T2IGMs have enabled advancements in data augmentation for computer vision tasks, allowing the generation of synthetic training data for improved model performance.

\begin{figure*}[!t]
    \centering
    \includegraphics[width=1\linewidth]{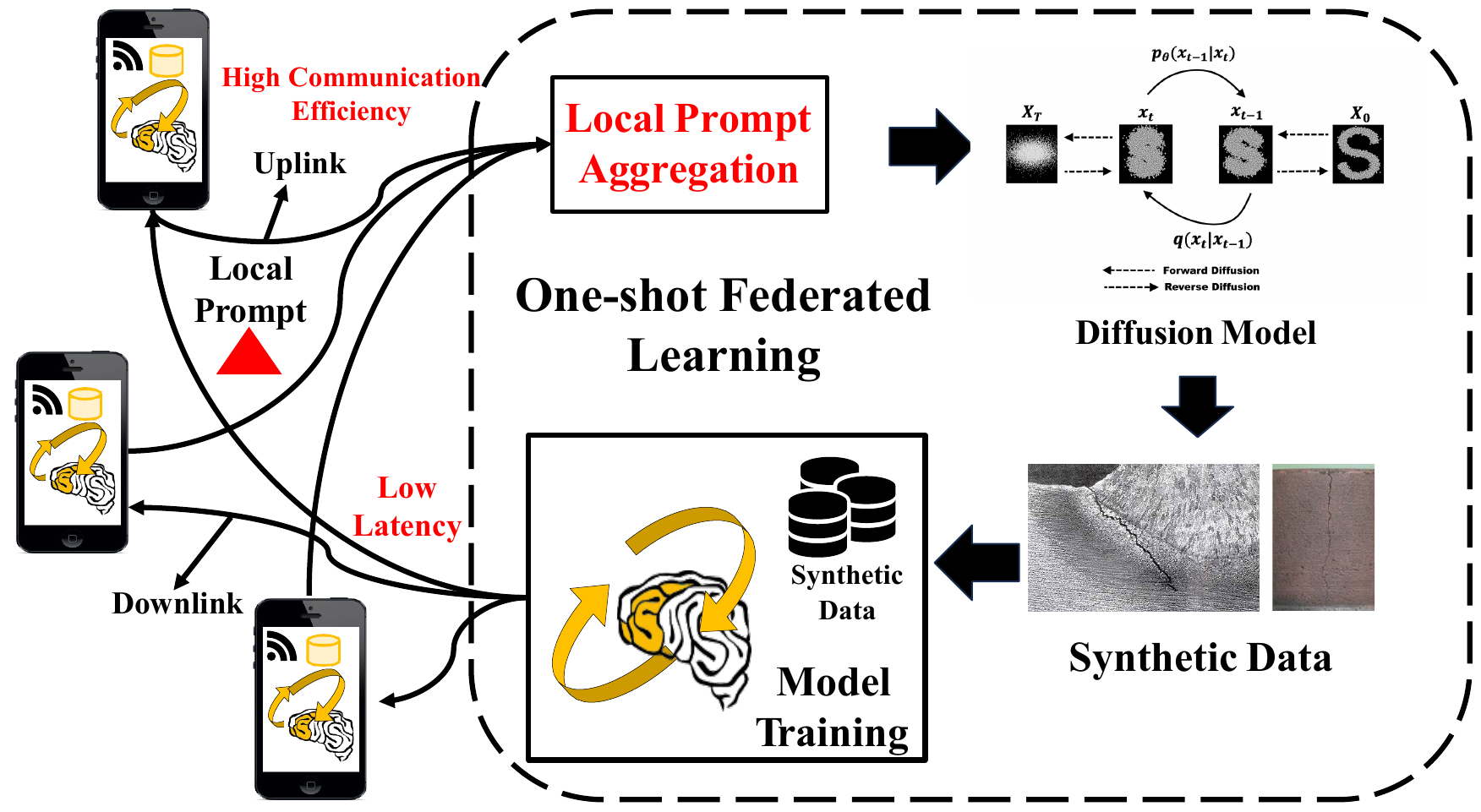}
    \caption{{An overview of the One-shot FGL system.} }
    \label{fig-1}
\end{figure*}

\section{Methodology}\label{sec-4}

\subsection{Overview of Federated Generative Learning Framework}
FGL, as shown in Figure \ref{fig-1}, is an innovative and emerging area of research that combines two powerful domains: FL and generative learning. FGL seeks to harness the potential of generative models to create synthetic data while preserving data privacy and decentralization principles in a federated environment. Here's an overview of the key components and concepts within Federated Generative Learning:
\begin{itemize}
    \item \textbf{Generative Models for Privacy-Preserving Data Synthesis:} FGL employs a generative model to create synthetic data samples on the server. This model uses generation prompts sent by each client to create synthetic data samples without centralizing them, preserving data privacy.
    \item \textbf{Edge Device Collaboration:} In contrast to conventional federated learning approaches, which typically involve the transmission of features, parameters, or gradients, our method transmits prompts that correlate with the private data stored on clients to the central server. This novel approach enhances privacy preservation and communication efficiency.
    \item  \textbf{Server-Side Aggregation:} The central server aggregates the prompts from edge devices, enhancing the global generative model. This one-shot aggregation might involve techniques to balance the contributions from each device and maintain model quality.
    \item  \textbf{Synthetic Data Generation:} Once a robust global generative model is established, it can be employed to generate synthetic data samples that closely resemble the data distribution present on edge devices.
\end{itemize}

\begin{figure*}[!t]
    \centering
    \includegraphics[width=1\linewidth]{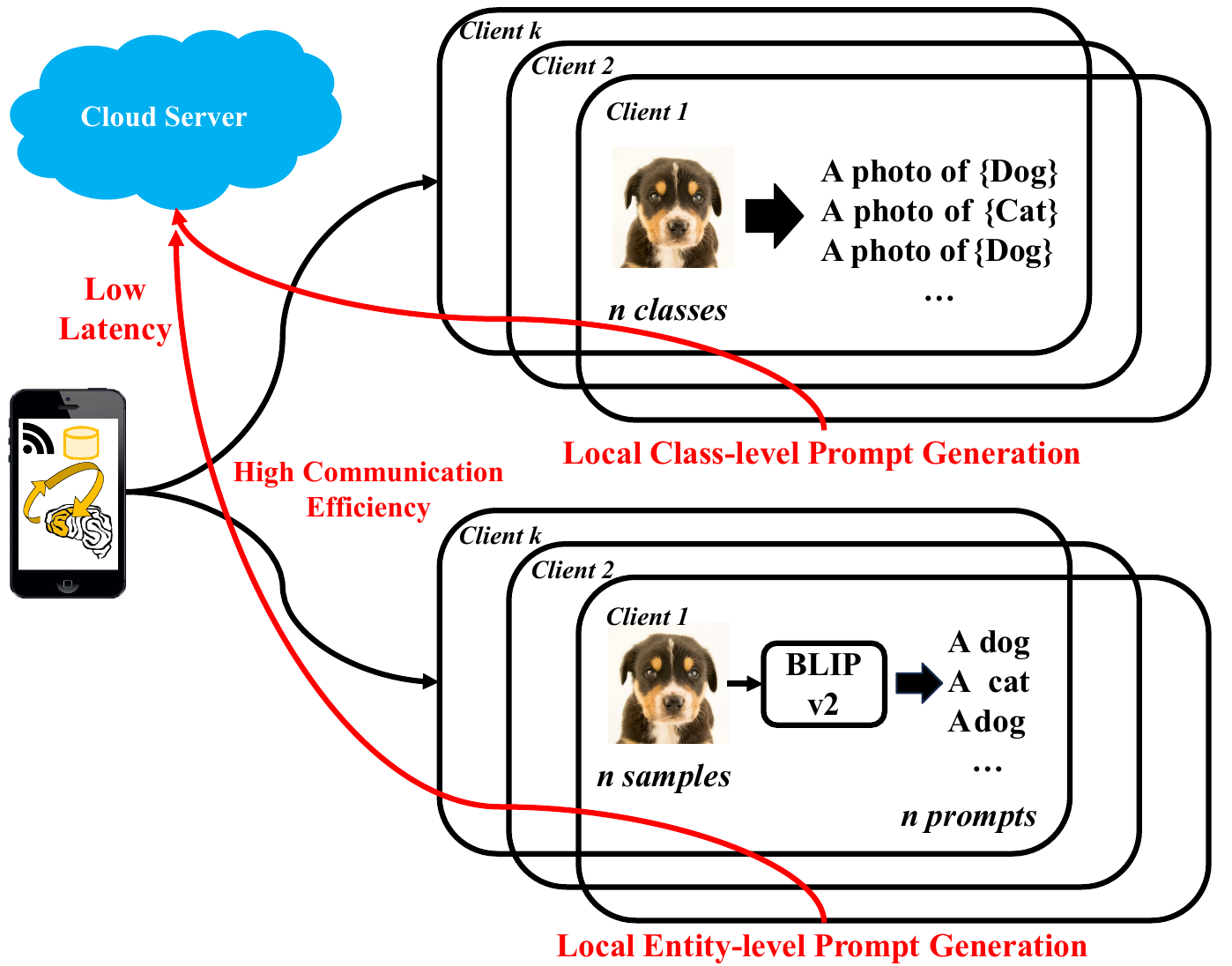}
    \caption{Local prompt generation strategies which are proposed by Zhang et al~\cite{zhang2023federated}.}
    \label{fig-2}
\end{figure*}

FGL offers numerous advantages, including enhanced data privacy, reduced communication overhead, and the ability to create synthetic data for data augmentation. It has applications in various fields, especially IoT, to create synthetic IoT sensor data for model training while protecting device privacy. It is worth noting that our method does not require multiple communications and only requires one-shot communication to complete model training, which greatly promotes the realization of green and sustainable IoT.

\subsection{Local Prompt Generation}
{In this paper, we follow \cite{zhang2023federated}, and introduce Local prompt generation, as shown in Figure \ref{fig-2}, which serves as a fundamental pillar in the groundbreaking paradigm of FGL.} FGL seamlessly merges the principles of FL and generative learning, offering a privacy-preserving and communication-efficient approach to data-driven tasks. At the heart of this innovative approach lies the concept of Local prompt generation, which empowers individual edge devices within a federated network to contribute their private data without compromising data privacy or centralized control.

{In this context, we utilize two distinct local prompt generation strategies~\cite{zhang2023federated}, both of which are visually depicted in Figure \ref{fig-2}.} The first strategy is designed to characterize classes within the local data, while the second focuses on characterizing individual entities present in the local datasets. These two prompt generation strategies share a common objective: to extract essential features from the local training data, facilitating the server-side generation model's ability to effectively collaborate in data synthesis. Importantly, it is essential to emphasize that these strategies do not entail the transmission of model updates or gradients. This unique characteristic substantially mitigates communication overhead, aligning with the core principles of efficient and privacy-preserving FGL.

\subsection{Training Data Synthesis}
\textcolor{black}{Training data synthesis constitutes a pivotal phase within the innovative paradigm of FGL. FGL seamlessly integrates the principles of FL and generative learning, with the overarching goal of revolutionizing the execution of data-driven tasks while upholding data privacy and decentralized control. In the FGL framework, training data synthesis assumes a transformative role, enabling individual edge devices to actively contribute to the development of a global generative model without exposing sensitive data.}

Training data synthesis in FGL commences with the generation of descriptive prompts by edge devices. These prompts encapsulate the essence of the local data they hold, serving as abstract yet informative representations. The prompts are designed to convey crucial insights about the data, such as class characteristics or entity attributes, without divulging raw data details. This synthesis of prompts aligns with the privacy-centric ethos of FL, as no actual data is shared during this process. 

Upon receiving all the prompts contributed by the edge devices, the central server embarks on the pivotal phase of data synthesis. {This process involves the generation of every training sample, denoted as $t_i$, by prompting the pre-trained generative models, \eg, Stable Diffusion model, with each prompt $p_i$. We follow \cite{zhang2023federated}, the sequence of actions can be summarized as follows:}

\begin{equation}
\boldsymbol{t}_i=G\left(\boldsymbol{z}_i, \boldsymbol{p}_i\right)=\sqrt{\beta} \sum_{t=1}^T \sqrt{1-\beta^t} \cdot \frac{1}{\sqrt{T}} \cdot G_{\boldsymbol{\vartheta}_t}\left(\boldsymbol{z}_i, \boldsymbol{p}_i\right),
\end{equation}
In this process, $z_i$ represents a randomly generated noise vector, $p_i$ signifies the prompt generated by edge devices, and $G_t(\cdot.\cdot)$ stands for the denoising network, which is parameterized with $t$ at the specific time step $t$. The hyperparameter $\beta$ plays a pivotal role in balancing image quality against diversity, while $T$ dictates the number of diffusion steps to be taken. This inference process follows an iterative pattern wherein the generated image undergoes denoising at each step before ultimately yielding the finalized synthetic training image. Subsequently, the central server orchestrates the generation of the synthetic training dataset, denoted as $\mathcal{S}$, comprising pairs of training samples and corresponding labels, represented as $\mathcal{S}=\left\{\left(s_i, y_i\right)\right\}_{i=1}^N$.

Crucially, training data synthesis in FGL operates without the exchange of model updates or gradients, a distinctive departure from traditional FL. Instead, it empowers edge devices to participate actively in the generative learning process without relinquishing control over their data. This approach not only enhances privacy preservation but also minimizes communication overhead, a critical consideration in resource-constrained environments like IoT.

\subsection{One-shot Learning}
One-shot learning streamlines the knowledge transfer process within FGL. In traditional FL, multiple rounds of model updates are exchanged between edge devices and the central server, often leading to substantial communication overhead. In contrast, one-shot learning condenses this into a singular communication round, dramatically reducing the amount of data transferred. The core objective of one-shot updating is to enrich the global generative model with collective insights from edge devices. These insights are encapsulated in the prompts generated by edge devices, which offer descriptive information about their local datasets. The central server leverages this valuable input to enhance the capabilities of the generative model. Following the acquisition of the synthetic training set denoted as $\mathcal{S}=\left\{\left(s_i, y_i\right)\right\}_{i=1}^N$, our subsequent step involves the joint training of the model on the central server. Specifically, we employ the AlexNet architecture as the model of choice and utilize the cross-entropy loss function. 

\begin{equation}
\mathcal{L}=-\frac{1}{N} \sum_{i=1}^N y_i \log \left(\hat{y}_i\right)+\left(1-y_i\right) \log \left(1-\hat{y}_i\right)
\end{equation}

The efficiency of one-shot updating contributes to the robustness and sustainability of FGL. It allows for the seamless integration of generative learning techniques into IoT and other data-sensitive domains, where resource-efficient machine learning is essential. Therefore, one-shot updating in FGL embodies the essence of efficient and privacy-conscious knowledge transfer. It empowers edge devices to contribute effectively to the generative learning process while preserving data privacy and minimizing communication overhead.

\subsection{Complexity Analysis}
Here, we focus on analyzing the algorithm complexity on the client side. It can be seen from the above algorithm that on the client side, the client needs to generate local prompts and upload the generated prompts. For the sake of simplicity, we let the complexity of generating a prompt be $C$ (depending on the specific generation algorithm), then the algorithm complexity of clients performing local prompt generation operations is $O(KCN)$. In addition, considering that the upload operation can be completed in one go, the complexity is $ O(N)$.

\SetCommentSty{mycommfont}
\begin{algorithm}[!t]
\DontPrintSemicolon
%   \KwInput{}
%   \KwOutput{Your output}
    \tcc{-- Local Class-level Prompt Generation --}
    \tcc{-- Step 1: Reading Local Data --}
    \Do{complete step 1}{
        read data class from local clients in the form of training samples\;
        get a new data class and add it to a list of unseen data class\;
    }
    % \tcc{----}
    % \;
    \While{local prompt generation}{    
    \tcc{-- Step 2: Marking the class --}
     Client to mark the number of classes\;
        \tcc{-- Step 3: Generating the prompt --}
        Client uses natural language to describe the tagged class\;
        }
    \tcc{-- Step 4: Uploading generated prompts --}
    upload the generated prompts to the server\;
    
    \tcc{-- Local Entity-leve Prompt Generation --}
    \tcc{-- Step 1: Reading Local Data --}
    \Do{complete step 1}{
        read data from local clients in the form of training samples\;
        get a new data  and add it to a list of unseen data \;
    }
    % \tcc{----}
    % \;
    \While{local prompt generation}{    
    \tcc{-- Step 2: Leveraging the model --}
     Client uses BLIPv2 model to exact data information\;
        \tcc{-- Step 3: Generating the prompt --}
        Client uses BLIPv2 model to describe the tagged sample\;
        }
    \tcc{-- Step 4: Uploading generated prompts --}
    upload the generated prompts to the server\;
    % \tcc{----}
\caption{Description of the steps of the local prompt generation algorithm.}
\label{alg-1}
\end{algorithm}

\SetCommentSty{mycommfont}
\begin{algorithm}[!t]
\DontPrintSemicolon
%   \KwInput{}
%   \KwOutput{Your output}
    \tcc{-- FGL Algorithm --}
    \tcc{-- Step 1: Initialize the Network --}
    prompts\_list = []\;
    global\_model = $\omega_0$\;
    \tcc{-- Step 2: Local Prompt Generation --}
    \For{client in all\_clients}{
    Uses class-level prompts or instance-level prompts via Algo. \ref{alg-1}\;
    prompts\_client = Algorithm\_1(client)\;
    prompts\_list.append(prompt)\;
    \tcc{-- Step 3: Local Prompt Aggregation --}
    prompts = prompts\_aggregation(prompts\_list)\;
    \tcc{-- Step 4: Training Data Synthesis --}
    synthetic\_dataloader = generative\_model(prompts, noises)\;
     \tcc{-- Step 5: One-shot Learniing --}
    Trains the global model at server side\;
     server\_update(synthetic\_dataloader, global\_model)\;
    }
\caption{Federated generative learning algorithm.}
\label{alg-2}
\end{algorithm}

\section{Experomental Results} \label{sec-5}

\subsection{Experiment Setup}\label{sec-6-1}
The experiments are conducted within a uniform computing environment, comprising Linux Ubuntu 10.04, an Intel i5-4210M CPU, 32GB RAM, and a 1024GB SSD. The implementation is facilitated through the utilization of the Pytorch and Foundation Model libraries. 

\noindent \textbf{Datasets.} We adopt three image datasets for evaluations, \ie, \textsf{Fashion-MNIST}~\cite{xiao2017fashion}, \textsf{CIFAR-10}~\cite{krizhevsky2009learning}, and \textsf{CIFAR-100}~\cite{krizhevsky2009learning}. The datasets cover different attributes, dimensions, and a number of categories, allowing us to explore the poisoning effectiveness of FGL.

\noindent \textbf{Fashion-MNIST.} Fashion-MNIST is a popular dataset in the field of machine learning and computer vision. It serves as an alternative to the traditional MNIST dataset but offers more challenging and diverse tasks. F-MNIST consists of 28x28 grayscale images of fashion items, such as clothing and accessories, categorized into ten different classes. These classes include items like T-shirts, dresses, sneakers, handbags, and more. F-MNIST is often used for benchmarking and evaluating machine learning algorithms and models, especially for tasks related to image classification, feature extraction, and deep learning.

\noindent \textbf{CIFAR-10.} The CIFAR-10 dataset is another well-known dataset in the field of computer vision. It contains 60,000 color images, divided into ten classes, with each class representing a different object category. Each image in CIFAR-10 has a resolution of 32x32 pixels and is labeled with one of the following categories: airplane, automobile, bird, cat, deer, dog, frog, horse, ship, or truck. CIFAR-10 is widely used for research in image classification, object recognition, and deep learning due to its diversity and complexity.

\noindent \textbf{CIFAR-100.} CIFAR-100 is an extension of the CIFAR-10 dataset and is designed for more fine-grained image classification tasks. It also consists of 60,000 images, but this time they are divided into 100 different classes. Each class represents a more specific object category, making CIFAR-100 suitable for tasks where distinguishing between closely related objects is required. The dataset includes objects like different types of birds, insects, and vehicles. CIFAR-100 is commonly used in research for tasks involving fine-grained object recognition, hierarchical classification, and model evaluation.

\noindent \textbf{Data Partition.} In FGL, data partitioning plays a pivotal role in shaping the training process. Setting up IID (Independent and Identically Distributed) data involves ensuring that each edge device's data is representative of the entire dataset and follows the same data distribution. To achieve this, the dataset is divided equally or approximately equally among the edge devices, ensuring that each device receives a balanced representation of data classes and characteristics. On the other hand, setting up non-IID (Non-Independent and Non-Identically Distributed) data in FGL involves recognizing the diversity of data among edge devices. Each device may possess a unique data distribution (we employ the Dirichlet distribution~\cite{li2022federated}), varying class proportions, or distinct data characteristics.

\noindent \textbf{Models.} For our experimental setup, we employ a straightforward deep learning model, specifically a Convolutional Neural Network (CNN) comprising two convolutional layers, succeeded by a single fully connected layer. This architecture is utilized for classification tasks involving the F-MNIST datasets. In contrast, for classification tasks related to the CIFAR-10 and CIFAR-100 datasets, we utilize the AlexNet~\cite{iandola2016squeezenet} model.

\noindent \textbf{Hyperparameters.} We configure our experiments with the following parameters: $m=100$ clients, client participation proportion $\mathcal{P}=0.1$, $T=250$ training rounds, $E=5$ local training epochs, learning rate $\eta = 0.001$, mini-batch size $B=32$, and $N_s=10,000$ training samples on the server-side. The symmetric quantization method employs an optimal threshold of $\alpha=0.5$. In the context of FGL, we adopt a hyperparameter value of $\beta =0.95$ . Additionally, our model optimization utilizes SGD with momentum applied to mini-batches

\noindent \textbf{Baselines.}  In our experimental setup, we establish baseline methods, specifically the widely adopted FedAvg approach proposed by McMahan et al. \cite{mcmahan2017communication}, and traditional centralized training for comparison. In this scenario, we consider a total of 5 clients, all actively participating in the communication process. For both centralized training and federated learning, we configure the local learning rate to be 0.01 and employ the SGD optimizer with a momentum of 0.9. During the FedAvg training phase, individual clients conduct local updates for 5 epochs, and the communication round is set to span 200 iterations. On the other hand, centralized training entails a total of 120 rounds of iterations as part of its training regimen.

\subsection{Numerical Results}
% \usepackage{booktabs}
% \usepackage{multirow}
% \usepackage{graphicx}
\begin{table}[!t]
\caption{System performance under the IID setting.}
\label{tab-2}
\resizebox{1\textwidth}{!}{%
\begin{tabular}{@{}lccccccccc@{}}
\toprule
\multicolumn{1}{c}{\multirow{2}{*}{Datasets}} & \multicolumn{3}{c}{F-MNIST} & \multicolumn{3}{c}{CIFAR-10} & \multicolumn{3}{c}{CIFAR-100} \\ \cmidrule(l){2-10} 
\multicolumn{1}{c}{} & \begin{tabular}[c]{@{}c@{}}Train \\ acc (\%)\end{tabular} & \begin{tabular}[c]{@{}c@{}}Test\\ acc (\%)\end{tabular} & Acc $\uparrow$& \begin{tabular}[c]{@{}c@{}}Train \\ acc (\%)\end{tabular} & \begin{tabular}[c]{@{}c@{}}Test\\ acc (\%)\end{tabular} & Acc $\uparrow$ & \begin{tabular}[c]{@{}c@{}}Train \\ acc (\%)\end{tabular} & \begin{tabular}[c]{@{}c@{}}Test\\ acc (\%)\end{tabular} & Acc $\uparrow$ \\ \cmidrule(r){1-10}
Centralized &96.47  &95.42 &95.42  &74.65  &67.45  &67.45  &45.65  &34.15  &34.15  \\
FedAvg~\cite{mcmahan2017communication} &95.10 &94.88 &94.88  &72.99  &64.20  &64.20  &42.85  &32.11 &32.11  \\
FGL (Ours) &98.58  &98.26  &\textbf{98.26}  &77.84  &70.85  &\textbf{70.85}  &58.67  &52.41  &\textbf{52.41}  \\ \bottomrule
\end{tabular}%
}
\end{table}

\begin{table}[!t]
\caption{System performance under the non-IID setting.}
\label{tab-3}
\resizebox{1\textwidth}{!}{%
\begin{tabular}{@{}lccccccccc@{}}
\toprule
\multicolumn{1}{c}{\multirow{2}{*}{Datasets}} & \multicolumn{3}{c}{F-MNIST} & \multicolumn{3}{c}{CIFAR-10} & \multicolumn{3}{c}{CIFAR-100} \\ \cmidrule(l){2-10} 
\multicolumn{1}{c}{} & \begin{tabular}[c]{@{}c@{}}Train \\ acc (\%)\end{tabular} & \begin{tabular}[c]{@{}c@{}}Test\\ acc (\%)\end{tabular} & Acc $\uparrow$& \begin{tabular}[c]{@{}c@{}}Train \\ acc (\%)\end{tabular} & \begin{tabular}[c]{@{}c@{}}Test\\ acc (\%)\end{tabular} & Acc $\uparrow$ & \begin{tabular}[c]{@{}c@{}}Train \\ acc (\%)\end{tabular} & \begin{tabular}[c]{@{}c@{}}Test\\ acc (\%)\end{tabular} & Acc $\uparrow$ \\ \cmidrule(r){1-10}
Centralized &96.47  &95.42 &95.42  &74.65  &67.45  &67.45  &45.65  &34.15  &34.15  \\
FedAvg~\cite{mcmahan2017communication} &89.24 &82.54 &82.54  &61.25  &53.61  &53.61  &38.98  &28.88 &28.88  \\
FGL (Ours) &97.24  &95.47  &\textbf{95.47}  &67.91  &64.33  &\textbf{64.33}  &41.25  &38.36  &\textbf{38.36}  \\ \bottomrule
\end{tabular}%
}
\end{table}

\noindent \textbf{Performance Evaluation.} In this section, we first evaluate the system performance of the proposed FGL under IID and non-IID data settings. As shown in Table \ref{tab-2} and Table \ref{tab-3}, we can find that the system performance of FGL on the F-MNIST and CIFAR-10 data sets is better than the baseline schemes. Specifically, on the CIFAR-10 dataset, the performance of FGL is 5.67\% higher than that of FedAvg. In addition, we can also find that the system performance of FGL is comparable to the baselines on the CIFAR-100 data set. After thorough research and reflection, we can get the following lessons: 

\begin{mybox}
\paragraphb{(Takeaway-1)} 
1) FGL's system performance is generally better than the baseline and does not require multiple rounds of communication, thanks to the powerful learning ability of the basic model. 2) Non-IID data setting does not significantly affect the system performance of FGL. 3) The OSFL method does not affect system performance.
\end{mybox}

\begin{figure}[!t]
    \centering
    \includegraphics[width=0.5\linewidth]{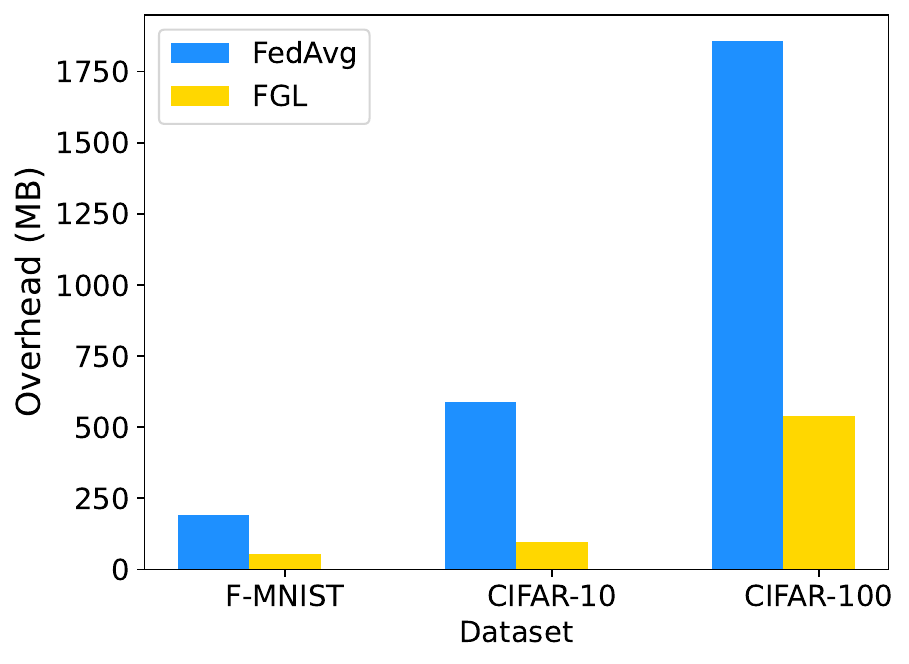}
    \caption{Communication efficiency evaluation on non-IID setting.}
    \label{fig-4}
\end{figure}

\begin{figure}[!t]
    \centering
    \includegraphics[width=0.5\linewidth]{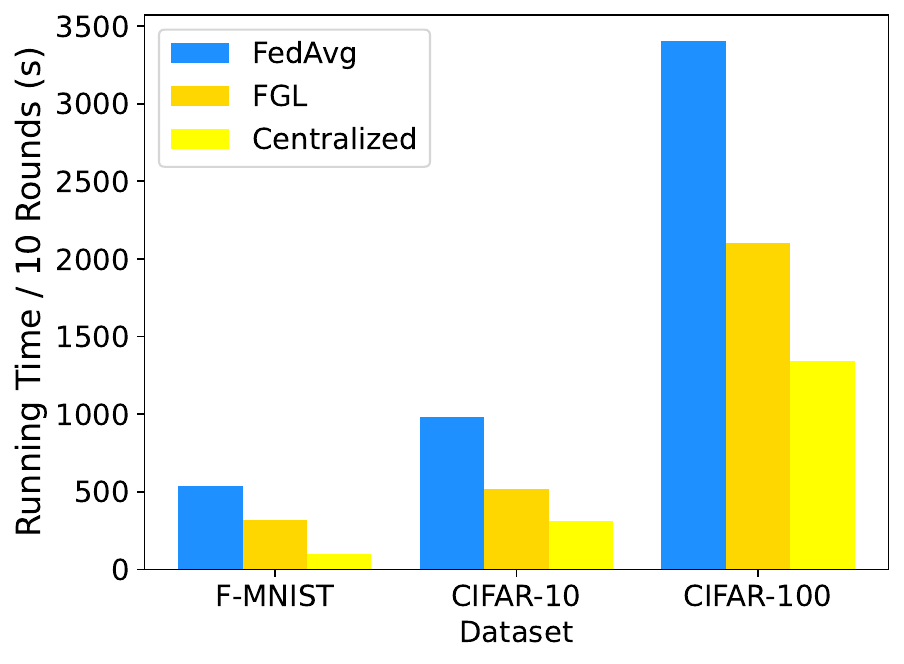}
    \caption{Running time reults on non-IID setting.}
    \label{fig-6}
\end{figure}

\noindent \textbf{Communication Efficiency Evaluation.} Second, we aim to explore the communication overhead between FGL and baselines. Since FGL utilizes one shot learning and updating methods, we expect that the communication overhead of FGL is much smaller than that of baselines. As shown in Figures \ref{fig-3} and \ref{fig-4}, we find that the communication overhead of FGL under IID and non-IID settings is approximately 1/5 of the FedAvg scheme. Note that since centralized training does not involve communication protocols it is not compared here. Furthermore, we note that the running time of FGL is approximately 1/2 of the baseline solutions, as shown in Figure \ref{fig-5} and Figure \ref{fig-6}. This means that FGL has significant advantages in communication efficiency and operating efficiency, and is especially suitable for resource-constrained IoT environments. After thorough research and reflection, we can get the following lessons: 

\begin{figure}[!t]
    \centering
    \includegraphics[width=0.5\linewidth]{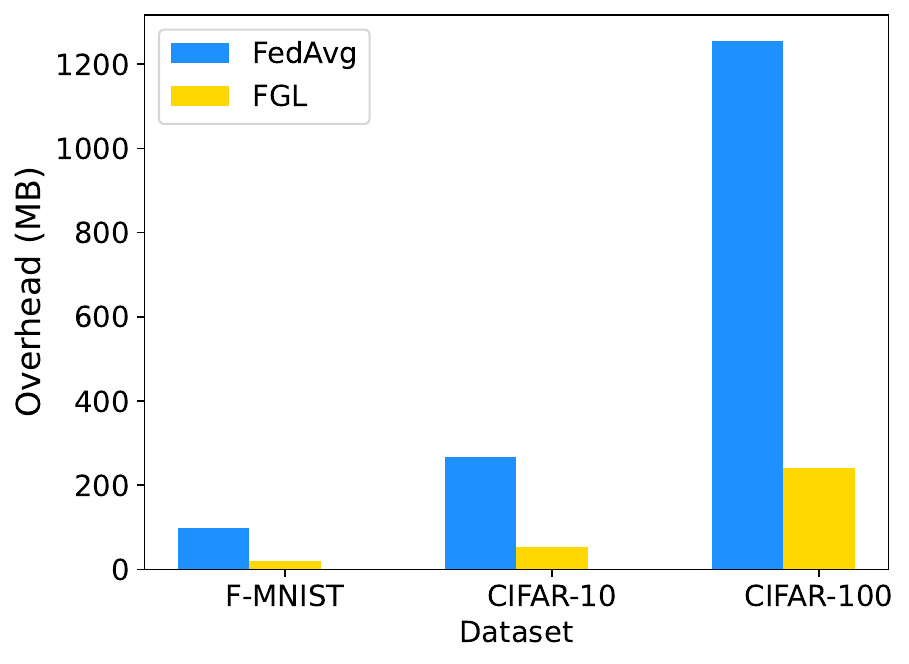}
    \caption{Communication efficiency evaluation on IID setting.}
    \label{fig-3}
\end{figure}

\begin{figure}[!t]
    \centering
    \includegraphics[width=0.5\linewidth]{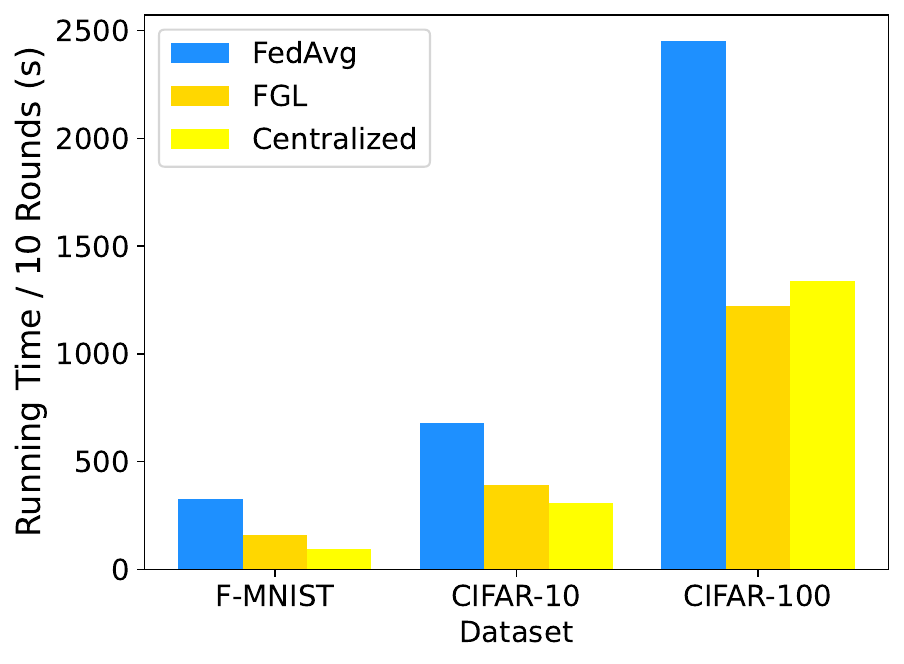}
    \caption{Running time reults on IID setting.}
    \label{fig-5}
\end{figure}

\begin{mybox}
\paragraphb{(Takeaway-2)} 
1) FGL benefits from the one-shot learning and updating method, which greatly reduces its communication overhead. 2) Since FGL no longer requires local training, its running efficiency has also been accelerated. 3) In resource-constrained IoT environments, FGL can achieve a perfect trade-off between model utility and resource energy consumption.
\end{mybox}

\noindent \textbf{Hyperparameter Evaluation under Number of Synthetic Data.} Third, we explore the impact of different amounts of synthetic data on system performance. Specifically, we set the number of synthetic data to $N=\{2k, 5k, 10k\}$, and record the system performance of FGL in these three cases. Table \ref{tab-4} reports the performance results of FGL, and we can find that as the amount of synthetic data increases, the performance of FGL continues to improve. However, this phenomenon cannot continue indefinitely. When the performance limit is reached, the increase in the amount of synthetic data is no longer beneficial to model training. After thorough research and reflection, we can get the following lessons:

\begin{table}[!t]
\caption{System performance under the number of synthetic data.}
\label{tab-4}
\resizebox{1\textwidth}{!}{%
\begin{tabular}{@{}lccccccccc@{}}
\toprule
\multicolumn{1}{c}{\multirow{2}{*}{\# of Synthetic Data}} & \multicolumn{3}{c}{2k} & \multicolumn{3}{c}{5k} & \multicolumn{3}{c}{10k} \\ \cmidrule(l){2-10} 
\multicolumn{1}{c}{} & \begin{tabular}[c]{@{}c@{}}Train \\ acc (\%)\end{tabular} & \begin{tabular}[c]{@{}c@{}}Test\\ acc (\%)\end{tabular} & Acc $\uparrow$& \begin{tabular}[c]{@{}c@{}}Train \\ acc (\%)\end{tabular} & \begin{tabular}[c]{@{}c@{}}Test\\ acc (\%)\end{tabular} & Acc $\uparrow$ & \begin{tabular}[c]{@{}c@{}}Train \\ acc (\%)\end{tabular} & \begin{tabular}[c]{@{}c@{}}Test\\ acc (\%)\end{tabular} & Acc $\uparrow$ \\ \cmidrule(r){1-10}
FGL (Ours) &77.84  &70.85  &\textbf{70.85}  &85.44  &82.54  &\textbf{82.54}  &89.22  &87.24 &\textbf{87.24}   \\ \bottomrule
\end{tabular}%
}
\end{table}

\begin{mybox}
\paragraphb{(Takeaway-3)} 
1) The number of synthetic data has an impact on the performance of FGL, and within a certain range, the larger the number, the more beneficial it is for model training. 2) Increasing the amount of synthetic data is a simple and economical way to improve performance.
\end{mybox}

\noindent \textbf{Visual Analysis of FGL’s Generalization Ability.} \textcolor{black}{As shown in Fig. \ref{fig-5}, we visualized the loss landscape of FGL and naive FL and found that the FGL framework not only has a more balanced optimization curve, but also has a lower loss value and is easier to obtain the optimal value. This means that although FGL is one-shot communication, it still has good convergence and generalization capabilities.}

\noindent \textbf{Performance Evaluation under Different Prompt Methods.} Finally, we explore the impact of different prompt methods on FGL system performance. As mentioned above, this article designs two local prompt strategies. We evaluate these strategies by comparing the system performance under different prompt strategies. Table \ref{tab-5} reports the results. We found that using a more precise entity-level prompt strategy leads to higher accuracy, for example, reaching 68.62\% on CIFAR-10 compared with the accuracy of the class-level prompt strategy. This result clearly highlights the importance of considering more detailed entity-level information in prompt strategy design. However, the entity-level prompt policy involves specific real image information, which may cause potential privacy issues, and we leave it as future work.

\begin{figure}[!t]    
  \centering    
  \begin{minipage}[b]{0.49\textwidth}    
    \includegraphics[width=\textwidth]{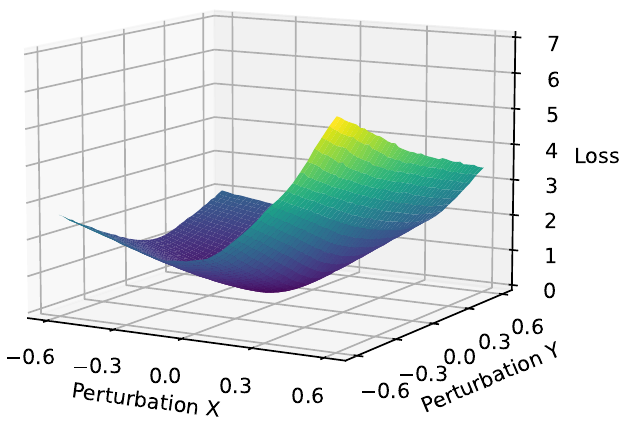}    
    \label{fig-5-1}    
  \end{minipage}%    
  \hfill    
  \begin{minipage}[b]{0.49\textwidth}    
    \includegraphics[width=\textwidth]{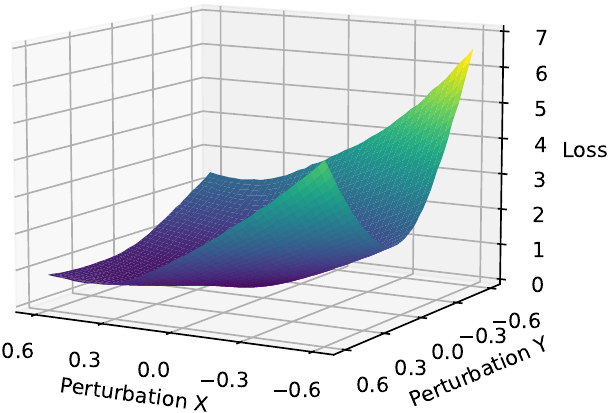}   
    \label{fig-5-2}    
  \end{minipage}%       
  \caption{(a) Visualization of loss landscape when applying FGL. (b) Visualization of the loss landscape when the FGL is not applied.}    
  \label{fig-5}   
\end{figure}

\begin{table}[!t]
\caption{System performance under different prompt methods.}
\label{tab-5}
\resizebox{1\textwidth}{!}{%
\begin{tabular}{@{}lccccccccc@{}}
\toprule
\multicolumn{1}{c}{\multirow{2}{*}{Datasets}} & \multicolumn{3}{c}{F-MNIST} & \multicolumn{3}{c}{CIFAR-10} & \multicolumn{3}{c}{CIFAR-100} \\ \cmidrule(l){2-10} 
\multicolumn{1}{c}{} & \begin{tabular}[c]{@{}c@{}}Train \\ acc (\%)\end{tabular} & \begin{tabular}[c]{@{}c@{}}Test\\ acc (\%)\end{tabular} & Acc $\uparrow$& \begin{tabular}[c]{@{}c@{}}Train \\ acc (\%)\end{tabular} & \begin{tabular}[c]{@{}c@{}}Test\\ acc (\%)\end{tabular} & Acc $\uparrow$ & \begin{tabular}[c]{@{}c@{}}Train \\ acc (\%)\end{tabular} & \begin{tabular}[c]{@{}c@{}}Test\\ acc (\%)\end{tabular} & Acc $\uparrow$ \\ \cmidrule(r){1-10}
Class-level &98.58  &98.26  &\textbf{98.26}  &77.84  &70.85 &\textbf{70.85}  &58.67  &52.41 &\textbf{52.41}  \\ 
Entity-level &99.62 &99.24  &\textbf{99.24}  &82.41 &78.62  &\textbf{78.62}  &61.36  &59.24  &\textbf{59.24}  \\ 
\bottomrule
\end{tabular}%
}
\end{table}

\begin{mybox}
\paragraphb{(Takeaway-4)} 
1) Both of the designed prompt strategies can effectively achieve the established goals. 2) The effect of entity-level prompt strategy is better than that of class-level prompt strategy. 3) Although the effect of the entity-level prompt policy is better than the class-level prompt policy, the entity-level prompt policy has certain privacy concerns.
\end{mybox}

\subsection{Discussion}
\textcolor{black}{Privacy has been a paramount concern in IoT ecosystems, and our One-shot FGL approach offers a pragmatic solution. By allowing edge devices to share abstract prompts instead of raw data, we strike a balance between data utility and privacy preservation. This design mitigates the risk of data breaches and unauthorized access while enabling the collaborative generation of synthetic data. The localization of data aggregation further fortifies privacy, as prompts, not individual data samples, are aggregated. The privacy-centric nature of FGL aligns with the ethical use of data and user expectations for data confidentiality in the IoT era.}

Sustainability is at the core of our approach. By reducing the need for extensive data transmission and centralization, FGL contributes to the ecological sustainability of IoT systems. The conservation of energy and bandwidth resources not only supports environmental goals but also enhances the operational longevity of IoT devices. FGL's ability to generate high-quality synthetic data from decentralized sources promotes sustainable practices by minimizing the environmental footprint of data transfer and storage. This integration of sustainability and privacy-consciousness positions FGL as a pioneering approach in building responsible and eco-friendly IoT ecosystems.

While our approach shows promise, several challenges and future directions warrant consideration. One key challenge is optimizing the generative model's performance while working with decentralized and diverse data sources. Developing advanced generative techniques that can adapt to various data distributions across edge devices is essential. Additionally, research in federated learning orchestration and model aggregation techniques needs further exploration to maximize the utility of FGL.

\textcolor{black}{Finally, it is crucial to highlight that the proposed framework exhibits flexibility and can be readily extended to various tasks. The adaptability stems from the ability to employ different generation models for prompt generation corresponding to diverse task data. For instance, the VisualGLM model can be employed to generate prompts and subsequently facilitate learning from video data. The scalability of the framework across different tasks is confirmed, underscoring the importance of task scalability in a learning framework.}

\section{Conclusions}\label{sec-6}

In this paper, we have introduced and delved into the concept of OSFL as an innovative approach aimed at addressing the dual imperatives of sustainability and privacy within the realm of the IoT. OSFL, characterized by its reduction of communication rounds and optimization for energy efficiency through generative learning, holds significant promise for reshaping the landscape of machine learning in IoT ecosystems. \textcolor{black}{Our research has successfully demonstrated the feasibility and effectiveness of OSFL in achieving sustainable IoT operations. By condensing the traditional federated learning process into a single communication round, OSFL substantially mitigates energy consumption, communication overhead, and latency. Moreover, the integration of generative learning techniques ensures the preservation of data privacy, even as knowledge is efficiently shared among IoT devices. OSFL stands as a novel and environmentally conscious paradigm that aligns with the global imperative to reduce the environmental footprint of technology while facilitating advanced machine learning capabilities.} 

Looking ahead, the potential applications of OSFL in green and sustainable IoT are expansive. From optimizing energy consumption in smart cities to enhancing environmental monitoring in remote regions, OSFL can empower IoT devices to learn and collaborate effectively while minimizing their environmental impact. Furthermore, OSFL's ability to balance sustainability and model performance opens doors to new opportunities in domains such as healthcare, agriculture, and industrial automation, where resource-efficient machine learning is crucial. In conclusion, OSFL not only underscores our commitment to responsible and sustainable technology development but also offers a transformative pathway to a greener and more intelligent IoT ecosystem.

\section*{Acknowledgment} The authors extend their appreciation to the Deputyship for Research and Innovation, Ministry of Education in Saudi Arabia, for funding this research through (IFKSURC-1-0306). Here, special thanks to Dr. Jie Zhang from ETH for the open-source federated generative learning code, which is the core basis of the work in this paper. We borrowed his code, and method as the cornerstone to develop the proposed green and, sustainable FL framework.

\printcredits

\bibliographystyle{cas-model2-names}

\bibliography{refs}

\end{document}